\documentclass[12pt,preprint]{aastex}
\def\degpoint{\ifmmode ^{\rm{o}}\!. \else $^{\rm{o}}\!.$\fi}

\newcommand{\ms}{\mbox{m\,s$^{-1}$}}

\newcommand{\Mjup}{\mbox{M$_{\rm Jup}$}}

\newcommand{\ltsimeq}{\raisebox{-0.6ex}{$\,\stackrel
         {\raisebox{-.2ex}{$\textstyle <$}}{\sim}\,$}}

\begin{document}

\title{Observing Strategies for the Detection of Jupiter Analogs }

\author{Robert A.~Wittenmyer\altaffilmark{1}, 
C.G.~Tinney\altaffilmark{1}, J.~Horner\altaffilmark{1}, 
R.P.~Butler\altaffilmark{2}, H.R.A.~Jones\altaffilmark{3}, 
S.J.~O'Toole\altaffilmark{4}, J.~Bailey\altaffilmark{1}, 
B.D.~Carter\altaffilmark{5}, G.S.~Salter\altaffilmark{1}, 
D.~Wright\altaffilmark{1} }
\altaffiltext{1}{Department of Astrophysics, School of Physics, Faculty 
of Science, The University of New South Wales, Sydney, NSW 2052, Australia}
\altaffiltext{2}{Department of Terrestrial Magnetism, Carnegie 
Institution of Washington, 5241 Broad Branch Road, NW, Washington, DC 
20015-1305, USA}
\altaffiltext{3}{University of Hertfordshire, Centre for Astrophysics 
Research, Science and Technology Research Institute, College Lane, AL10 
9AB, Hatfield, UK}
\altaffiltext{4}{Australian Astronomical Observatory, PO Box 915,
North Ryde, NSW 1670, Australia}
\altaffiltext{5}{Faculty of Sciences, University of Southern Queensland, 
Toowoomba, Queensland 4350, Australia}

\email{
rob@phys.unsw.edu.au}

\shortauthors{Wittenmyer et al.}

\begin{abstract}

\noindent To understand the frequency, and thus the formation and 
evolution, of planetary systems like our own solar system, it is 
critical to detect Jupiter-like planets in Jupiter-like orbits.  For 
long-term radial-velocity monitoring, it is useful to estimate the 
observational effort required to reliably detect such objects, 
particularly in light of severe competition for limited telescope time.  
We perform detailed simulations of observational campaigns, maximizing 
the realism of the sampling of a set of simulated observations.  We then 
compute the detection limits for each campaign to quantify the effect of 
increasing the number of observational epochs and varying their time 
coverage.  We show that once there is sufficient time baseline to detect 
a given orbital period, it becomes less effective to add further time 
coverage -- rather, the detectability of a planet scales roughly as the 
square root of the number of observations, independently of the number 
of orbital cycles included in the data string.  We also show that no 
noise floor is reached, with a continuing improvement in detectability 
at the maximum number of observations $N=500$ tested here.

\end{abstract} 

\keywords{planetary systems -- techniques: radial velocities }

\section{Introduction}

As planet search efforts mature, techniques improve, and ever-larger 
data sets are obtained, the detection of planets like our own Jupiter is 
becoming more tractable.  These ``Jupiter analogs'' are critically 
important for understanding the frequency of planetary systems with 
architectures similar to our own solar system \citep[e.g.][]{cumming08, 
gould10, jupiters}.  Jupiter analogs are interesting and important for a 
number of reasons.  Firstly, most exoplanetary systems found to date are 
markedly different from our own Solar system, featuring Jupiter-mass 
planets on short-period orbits more reminiscent of the terrestrial 
planets.  While such systems are fascinating in their own right, it is 
clearly important to search for systems that more closely resemble our 
own.  This is particularly true looking forward, when we consider the 
ongoing efforts of planet-search programs to find Earth-like planets 
that could potentially host life.

Over the years, the presence of Jupiter-analogs has been invoked as a 
key factor in determining the habitability of Earth-like planets 
\citep{ward00, greaves04}.  Although the idea that Jupiter-analogs are 
required to shield terrestrial planets from impacts has been 
conclusively dismantled \citep[e.g.][]{hj08, horner10, horner12}, the 
presence of such planets could still prove to be an important mechanism 
that drives the delivery of water to planets that might otherwise have 
formed as dry, lifeless husks.  For an in-depth discussion of this idea, 
we direct the interested reader to p.285-6 of Horner \& Jones (2010), 
and references therein.

Regardless of their role in ensuring (or threatening) the habitability 
of telluric worlds, the search for Jupiter-analogs will provide a key 
datum for models of planetary formation and evolution - attempting to 
answer the question ``how common are planetary systems like our own.'' 
To date, the most prolific means of detecting Jupiter-analog planets has 
been the radial-velocity method.  While direct imaging and microlensing 
have contributed to the pool of known Jupiter analogs 
\citep[e.g.][]{1207B, marois10, gaudi08, dong09}, those planets detected 
by radial velocity remain the most amenable for detailed orbital 
characterisation.  The radial-velocity technique is well-established, 
having been in use by several planet-search teams for nearly 20 years.  
A key disadvantage of searching for Jupiter analogs in this way is that 
one must observe for at least a full orbital cycle to properly constrain 
the period and amplitude of a planet candidate.

For this sort of work, there is no substitute for time.  The 
Anglo-Australian Planet Search (AAPS) has been in operation for nearly 
14 years, and has discovered some 40 planets 
\citep[e.g.][]{tinney01,butler01,jones06,jones10,tinney11}.  A key 
strength of this program is that the AAPS has used the same instrumental 
setup throughout its lifetime: the UCLES spectrograph on the 3.9m 
Anglo-Australian Telescope.  AAT+UCLES has enabled the AAPS to amass a 
database of velocities with precisions of 2-3 \ms\ for most of its 
target stars.  This is exactly the type of data set required to robustly 
detect the signals of Jupiter-like planets in Jupiter-like orbits.  The 
long-term velocity stability achieved by the AAPS has enabled the 
discovery of six such planets to date: HD\,30177b (Butler et al.~2006: 
$P=7.6$yr, M~sin~$i$=9.7\Mjup) HD\,160691c (McCarthy et al.~2004: 
$P=11.5$yr, M~sin~$i$=1.9\Mjup), GJ\,832b (Bailey et al.~2009: 
$P=9.4$yr, M~sin~$i$=0.6\Mjup), HD\,134987c (Jones et al.~2010: 
$P=13.7$yr, M~sin~$i$=0.8\Mjup), HD\,142c (Wittenmyer et al.~2012: 
$P=16.4$yr, M~sin~$i$=5.3\Mjup), and HD\,4732c (Sato et al.~2013: 
$P=7.5$yr, M~sin~$i$=2.4\Mjup).

Given the significant and ever-tightening constraints on large-telescope 
time, it is prudent to optimise the observing strategy used to detect 
particular classes of planets \citep{ford08,bottom13}.  The AAPS, in 
recognition of its primary strength in detecting Jupiter-like planets, 
has recently adjusted its target list and observing strategy to be able 
to make meaningful, quantitative statements on the frequency of these 
types of planets \citep{jupiters}.  In this paper, we examine the impact 
of various observing strategies on the detectability of Jupiter analogs.  
Specifically, we ask what number and duration of observations would most 
efficiently enable the detection of planets with velocity amplitudes 
similar to that of Jupiter ($K=12$\,\ms).  We construct simulated data 
sets which build on the extant AAPS data (Section 2), and we use these 
results (Section 3) to draw general conclusions (Section 4) on the 
optimal approach for detecting Jupiter analogs in the face of extremely 
limited telescope time.

\section{Numerical Methods}

\subsection{Constructing the simulated data sets}

From the main AAPS sample of $\sim$250 stars, we wish to define a 
subsample of stars which have both low intrinsic stellar activity and a 
sufficient observational baseline to be useful for the detection of 
Jupiter analogs.  Following our previous work on this subject 
\citep{jupiters}, we chose those stars which have, at present, at least 
30 epochs over a time baseline of at least 3000 days.  We then excluded 
those stars which did not satisfy the criterion $(RMS/\sqrt{N}) \ltsimeq 
1$, an empirical relation derived from the AAPS data examined in 
\citet{jupiters} which can be used as a simple estimate of whether the 
12\,\ms\ velocity amplitude of Jupiter would be detectable at 99\% 
confidence.  After this cut, 103 stars remained, and all further 
analysis was done using those data.

\begin{figure}
\plotone{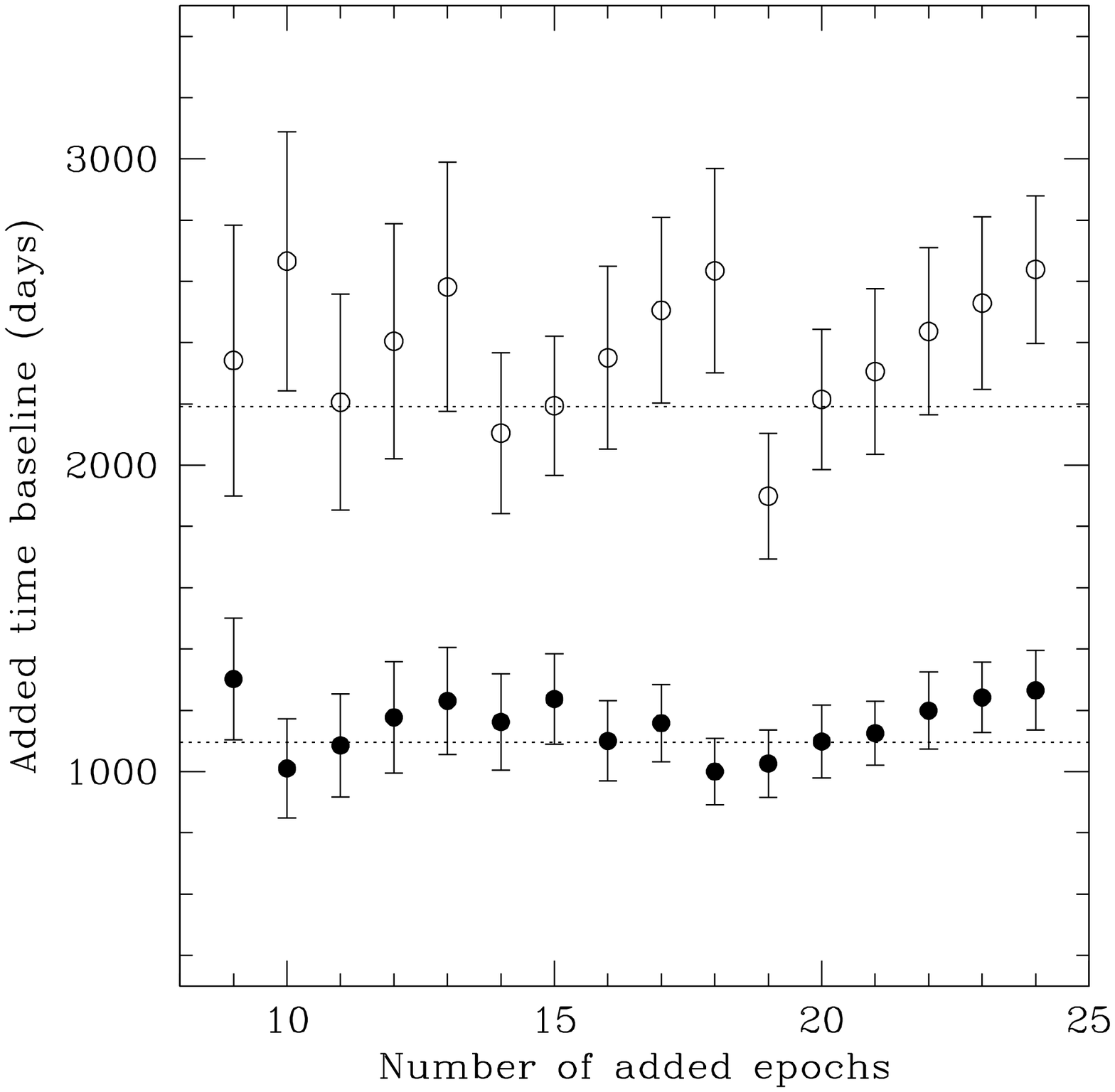}
\caption{Duration of added simulated observations tested here.  Each 
point represents the mean of the added time coverage over the 103 stars 
considered.  Dashed lines are at 3 and 6 years, which are the nominal 
durations of the added observations.  Filled circles -- 3 added years; 
open circles -- 6 added years. }
\label{duration}
\end{figure}

The epochs of real observational data are never purely random: stars can 
only be observed at night, and most targets are unobservable for a 
portion of the year.  When combined with the exigencies of telescope 
scheduling (planet-search programs are usually assigned time during 
bright lunations) real data can contain significant gaps.  To better 
simulate the sampling characteristics of real data, we made the 
following assumptions: (1) one observation in a 10-night block every 30 
days (bright-time scheduling), (2) the target is unobservable for four 
consecutive months every year, and (3) poor weather randomly prevents 
the observation 33\% of the time.  These conditions were selected for 
the following reasons: (1) planet-search programs are usually allocated 
time in bright lunations owing to the brightness of the targets, (2) for 
a mid-latitude site such as the Anglo-Australian Telescope (AAT), with 
planet-survey targets distributed randomly in right ascension, the 
average target is unobservable for four months in a year\footnote{Here 
we define ``unobservable'' to mean that the target spends less than one 
hour at an airmass less than 2.}, and (3) the long-term weather 
conditions at the AAT yield a 33\% loss rate (this ``weather allowance'' 
is accounted for in the proposal process).  Using this procedure, we 
added simulated observations to the current AAPS data (as of 2012 July) 
until reaching a desired number of added data points (chosen to lie 
between 9 and 24\footnote{These values were chosen to reflect the 
minimum (1.5 epochs/yr in 6 years) and maximum (8 epochs/yr for 3 years) 
practical cadences for a typical shared large telescope}).  In the 
interest of informing strategic plans for planet-search programs, we 
investigated the effect of adding observations over two timescales: 
three and six years.  The frequency of observations (e.g.~monthly or 
alternate months) was adjusted so that the resulting total time baseline 
of new data remained approximately constant at three and six years.  
Figure~\ref{duration} shows the time baseline added using this approach, 
averaged over the 103 stars.  In total, 32 artificial data sets were 
created for each target considered here: $N=9-24$ points added over 3 
and 6 year periods.  The radial velocity of the simulated data point is 
drawn from a Gaussian distribution with a FWHM equal to the RMS of the 
existing data for that star.  The uncertainty on the simulated radial 
velocity is drawn at random (with replacement) from the uncertainties of 
the existing real data.  The result is a simulated future data set with 
the same properties as the original data.

\subsection{Assessing Detectabilities}

Throughout this work, we assessed the detectability of radial-velocity 
signals using the algorithm described extensively in our previous work 
\citep{limitspaper, jupiters}.  Here we only consider circular orbits 
(after e.g.~Cumming et al.~2008), as previous work has shown that 
circular-orbit detection limits are a good approximation to the result 
that would be obtained for planets with $e\ltsimeq$0.5 \citep{cumming09, 
foreverpaper, jupiters}.  In brief, for each of 100 trial periods, a 
Keplerian signal is added to the data, then a Lomb-Scargle periodogram 
\citep{lomb76, scargle82} is used to search for the injected signal.  
The radial velocity semi-amplitude $K$ of the artificial planet is 
increased until 100\% of the signals are recovered with at least 99.9\% 
significance at the correct period.  That value of $K$ is then the 
detection limit for that trial period.  As we are concerned with 
long-period planets, we only consider periods from 1000-5000 days, 
evenly distributed in log space.  Throughout this work, the figure of 
merit used to assess the impact of added data is the mean detection 
limit $K$ averaged over 100 trial periods in the range 1000-5000 days.

\section{Results}

\subsection{Duration of Observations}

\begin{figure}
\plotone{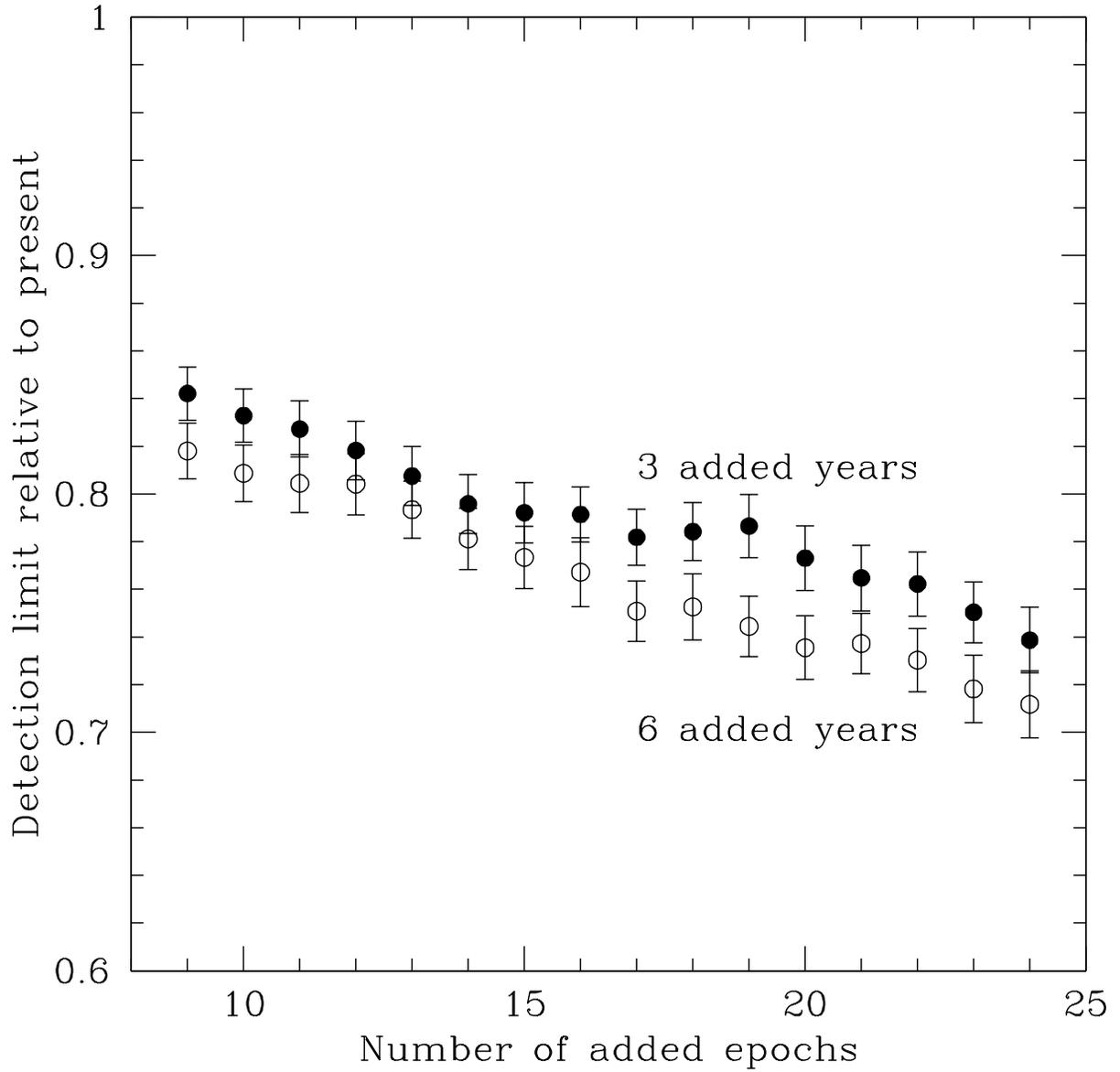}
\caption{Ratio of the mean detection limit $K$ achieved by adding 
simulated data as compared to the present value (averaged over 103 
stars).  Error bars represent the standard error of the mean ratios.  
Filled circles -- 3 added years; open circles -- 6 added years. }
\label{compare}
\end{figure}

The most obvious question here is: ``How does the number of added epochs 
affect the detectability of Jupiter analogs?'' A related issue is 
whether there is a point of diminishing return, where adding more data 
or more frequent observations does not make meaningful improvements in 
the overall detectability.  To determine the improvement to be had by 
various levels of observational effort, we also computed the detection 
limits for the 103-star sample exactly as above, but using only the 
existing AAPS data.  This gives a ``benchmark'' detectability for each 
star, against which we can then compare the results from the simulated 
observations on a star-by-star basis.  For each star, we divide the mean 
$K$ limit (averaged over periods from 1000-5000 days) by that obtained 
for the ``benchmark'' or present data set.  The grand means of these 
ratios (averaged over the 103 stars) are plotted in 
Figure~\ref{compare}.  From this figure, we can answer two questions: 
(1) What is the effect of adding epochs (within a fixed time), and (2) 
What is the impact of the observational baseline for a given number of 
epochs?  Our results show that, for the range of added epochs simulated 
here, there remains a linear relationship between the number of epochs 
and the achievable detection limit.  That is, the points shown in 
Figure~\ref{compare} do not level off, up to 24 added epochs.  On the 
second question, we see that there is no statistically significant 
difference between $N$ epochs added over 3 years and $N$ epochs added 
over 6 years, though this applies only for planets with $P<5000$ days 
(14 years) as this is the maximum period considered in the 
detection-limit calculation.

\subsection{Number of Observations}

The results of this experiment, in which we added simulated observations 
to real AAPS data, show that the amount of added time coverage is less 
important than the \textit{number} of added observations.  It is worth 
noting here that the selected stars all had at least 3000 days (8 years) 
of time coverage before adding new data.  Adding about 1000 or 2000 days 
(3 or 6 years) of new data thus has little effect on the detectability 
of Jupiter-like planets in the 1000-5000 day period range.  We also note 
that the \textit{total} number of epochs for each star varies greatly, 
with some stars at present having $\sim$35 epochs, and others having 
well over 100.

\begin{figure}
\plotone{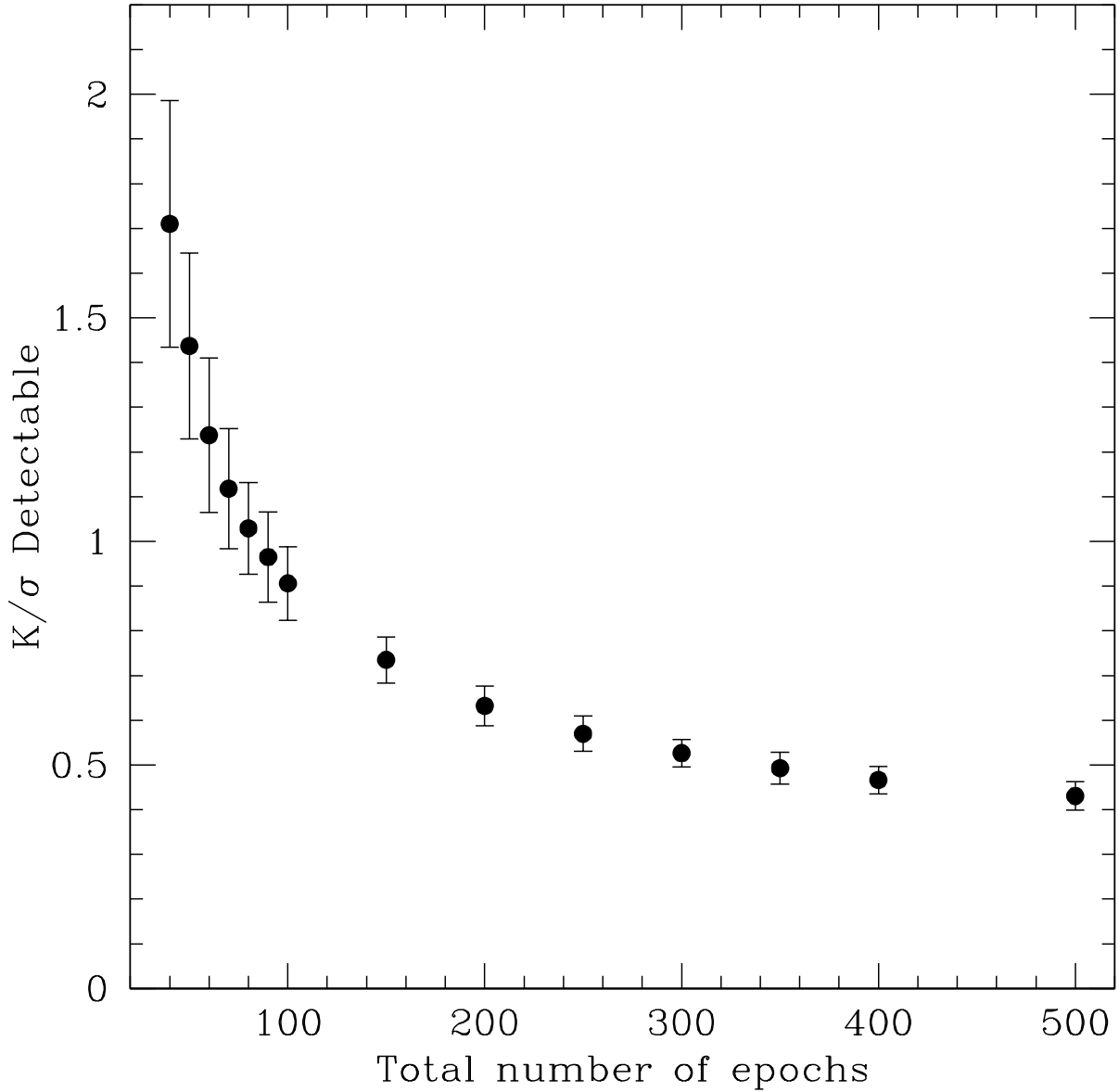}
\caption{Radial-velocity signal-to-noise $K/\sigma$ detectable in 
simulated data sets with various numbers of observations.  For each of 
100 data sets with a given $N$, we average (over 100 trial periods) the 
detection limit $K$ for which 99\% of injected planets were recovered, 
then divide by the total RMS of the input data.  Each point represents 
the grand mean of these mean $K$ values derived from the 100 data sets.  
Error bars are the standard deviation about that grand mean, also 
normalised by the total RMS of the input data.  These results show that 
a noise floor is reached for $N>250$, beyond which further observations 
do not improve the overall detection limit. }
\label{noisefloor}
\end{figure}

Since this could have biased the results shown in the 
Figure~\ref{compare}, we ran a second suite of simulations with slightly 
different initial conditions -- less ``realistic'' but more controlled.  
We created wholly artificial data sets with a fixed number of 
observations $N$, ranging from 40 to 500 epochs.  The time sampling was 
done exactly as described above, with the natural result that data sets 
with more observations had a longer timespan.  The average time coverage 
of the simulated data sets ranged from 7.2 years ($N=40$) to 92 years 
($N=500$).  We have chosen the minimum $N$ as 40 epochs, which yields a 
time baseline of 7.2$\pm$0.7 years, consistent with the initial 
selection of AAPS stars used in the first suite of simulations.  Also, 
in recognition of the fact that radial-velocity noise is almost 
certainly not Gaussian in nature \citep{otoole09a, otoole09b, taucet}, 
we chose instead to draw the value of the simulated data point (and its 
uncertainty) from a ``master'' set of 531 observations from a selection 
of known stable stars in the AAPS catalog.  Those velocities have an RMS 
of 3.0 \ms, which is typical of the long-term accuracy and precision 
achieved by AAT/UCLES over the lifetime of the AAPS.  In this way, we 
assure that the simulated data are purely noise, with the 
characteristics of velocity noise from actual observed stars.  The 
result is 100 simulated data sets each with $N$ epochs, for $N$ between 
40 and 500.  We computed the detection limits as described above, and 
calculated the mean velocity amplitude $K$ averaged over 100 trial 
periods for each data set, again averaging over periods from 1000-5000 
days.  Trial periods for which all simulated planets were not detected 
were ignored.  Table~\ref{fakedata} and Figure~\ref{noisefloor} show the 
results, where we have averaged the mean $K$ from each data set over all 
100 sets at each value of $N$.

\begin{figure}
\plotone{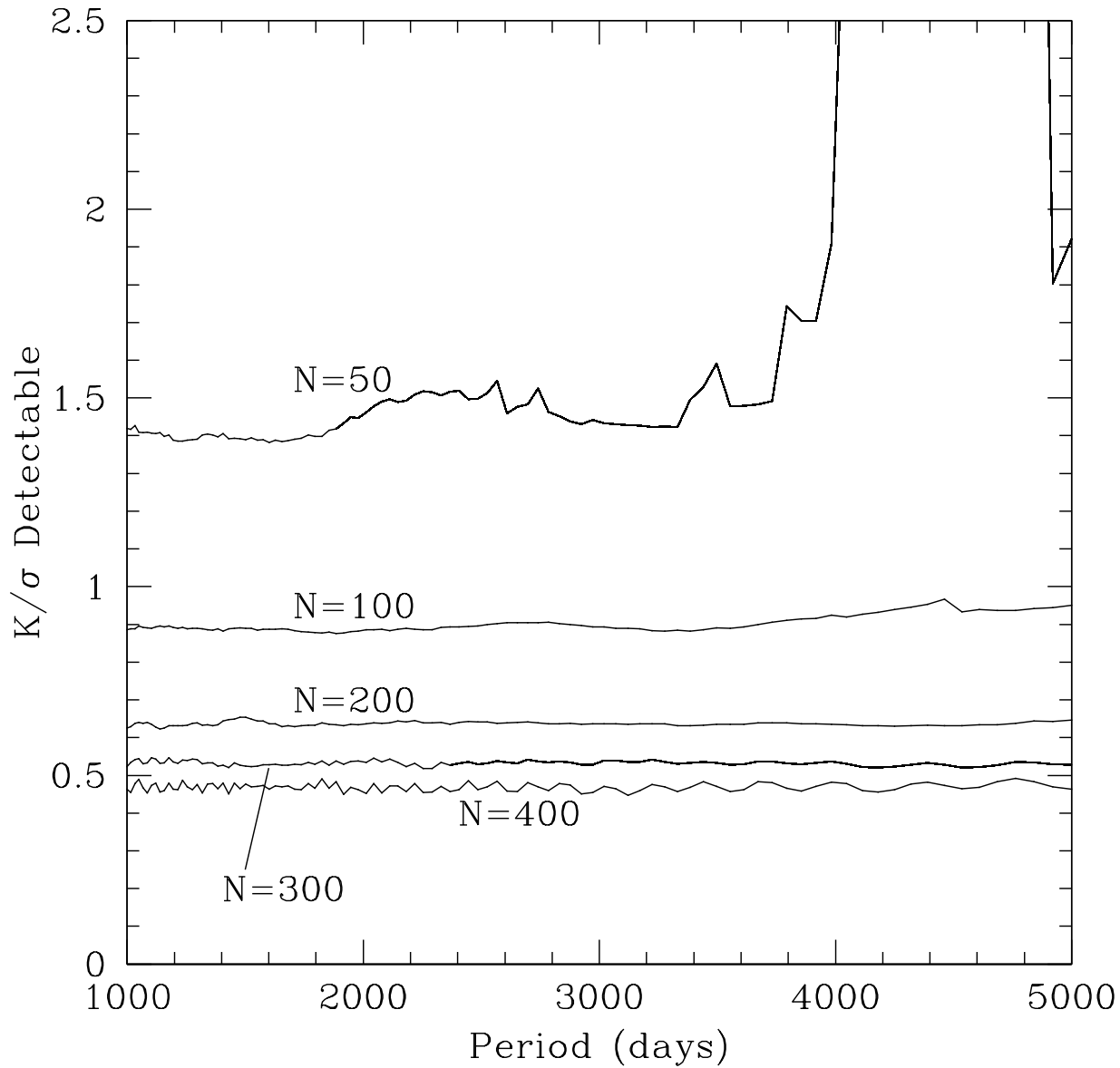}
\caption{Radial-velocity signal-to-noise $K/\sigma$ detectable in 
simulated data sets with various numbers of observations.  To show the 
dependence of our results on the orbital period, we have used the same 
data from Section 3.2 and Figure~\ref{noisefloor}, but averaged over the 
100 simulated data sets \textit{at each period}.  Each solid line thus 
represents the $K/\sigma$ at each period, averaged over the 100 distinct 
trials for each number of observations $N$.  The poor detectabilities at 
long periods seen for the $N=50$ case result from an insufficient time 
baseline. }
\label{perioddependence}
\end{figure}

We see from Table~\ref{fakedata} and Figure~\ref{noisefloor} that the 
detection limit $K$ scales approximately as $\sqrt{N}$, with the 
greatest marginal improvement occurring when the total duration extends 
beyond $\sim$4000 days (i.e. for $N=40-60$).  This is the expected 
result for data consisting of Gaussian (white) noise (e.g. Cochran \& 
Hatzes 1996); for the purposes of long-period planet detection, the 
radial-velocity noise distribution may be closer to Gaussian than is 
commonly expected.  Up to the limit of our tests, $N=500$, no noise 
floor is reached - the detectabilities continue to improve as 
$\sqrt{N}$.  We emphasize that this result applies for planets in the 
period range we have tested here: between 1000 and 5000 days.  
Obviously, continuing to observe a given star indefinitely would permit 
the detection of arbitrarily long periods shorter than the total 
duration of observations.  The period dependence of our results is shown 
in Figure~\ref{perioddependence}.  Obviously, for the $N=50$ case, where 
the total time coverage is only about 9 years (3300 days), periods 
substantially longer than this baseline are virtually undetectable.  
This artifact disappears once the time coverage exceeds the 5000-day 
maximum trial period, and we see no significant period dependence for 
the larger-$N$ trials.

The mean detection limit drops below the 3.0 \ms\ scatter of the input 
data after about 80 points (Figure~\ref{noisefloor}, and reaches 
$K/\sigma\,=$0.5 when 300-350 epochs are obtained.  Our result is 
applicable to other data sets with different noise properties -- the 
detection limit would then simply scale with the overall noise.  This 
can be seen from recent HARPS results, where the lowest-amplitude 
planetary signals are indeed at levels consistent with our results in 
Figure~\ref{noisefloor}.  For example, HD\,10180b has $K/\sigma\,=$0.62 
with $N=190$ measurements \citep{lovis11}.  Similarly, \citet{pepe11} 
report HD\,20794c with $K/\sigma\,=$0.68 and $N=187$ measurements.  
Because most radial-velocity planet search programs on shared large 
telescopes have similar sampling constraints, our results can thus be 
applied to other data sets by scaling the noise accordingly.

\section{Conclusions}

Taken together, the tests we have performed can inform the strategies 
used by long-running radial-velocity planet search programmes such as 
the AAPS, in order to make best use of limited telescope time to 
efficiently detect (or rule out) Jupiter-like planets.  We have shown 
that, once there is sufficient observational baseline to detect 
long-period Jupiter analogs, there is not much added benefit to 
extending that time base (except to detect ever-longer periods, e.g. 
Saturn analogs with 30-year periods).  That is, for a Keplerian velocity 
signal with a given orbital period and amplitude $K$, including more 
orbital cycles in the observed data does not have a significant effect 
on the detectability of that signal.  

We have also performed detailed simulations of radial-velocity 
observations, sampled according to a realistic schedule and with noise 
characteristics identical to actual data from the Anglo-Australian 
Planet Search.  Those simulations showed that the Keplerian velocity 
amplitude $K$ detectable from a data set scales approximately as 
$1/\sqrt{N}$, to the maximum $N=500$ tested.  That is, the 
radial-velocity noise may be closer to white than one might have 
expected.  

We note, however, that for extremely low-amplitude signals such as those 
detected for Alpha Centauri B \citep{dum12} and Tau Ceti \citep{taucet}, 
stellar noise at the $<1$ \ms\ level has a critical impact on the 
detectability of such small signals.  The red noise introduced by star 
spots, differential rotation, and convective blueshift is of an 
amplitude too small to be especially relevant to this work.  In this 
work, we have concentrated on the detectability of Jupiter-like planets 
with relatively large $K$ amplitudes of order 10 \ms.

We conclude that it remains worthwhile to continue radial-velocity 
observations of suitably stable stars to robustly detect or exclude 
long-period giant planets. However, the targets must be chosen carefully 
to ensure that the $K\sim$10-15 \ms\ signals of Jupiter-like planets are 
reliably detectable.  For ever-longer orbital periods, approaching 
Saturn-like orbits ($P\sim$30 yr), opportunities are now emerging to 
combine the complementary strengths of legacy radial-velocity data with 
rapidly improving direct-imaging technology from new instruments such as 
the Gemini Planet Imager.  The powerful combination of these two 
approaches will soon yield direct measurements of the occurrence rate of 
Jupiter analogs, and the first detailed characterisation of such objects 
orbiting nearby Sun-like stars.

\acknowledgements

We thank the anonymous referee for a timely and thoughtful report, which 
improved this manuscript.  This research has made use of NASA's 
Astrophysics Data System (ADS), and the SIMBAD database, operated at 
CDS, Strasbourg, France.  This research has also made use of the 
Exoplanet Orbit Database and the Exoplanet Data Explorer at 
exoplanets.org \citep{wright11}.



\begin{deluxetable}{lccc}
\tabletypesize{\scriptsize}
\tablecolumns{4}
\tablewidth{0pt}
\tablecaption{Detectabilities from Artificial Data Sets }
\tablehead{

\colhead{$N$} & \colhead{Mean $K$ (\ms)} & \colhead{$K_{i}/K_{i-1}$} & 
\colhead{$\sqrt{N_{i-1}/N_{i}}$}

}
\startdata
\label{fakedata}
40  & 5.11$\pm$0.82 & \nodata & \nodata \\
50  & 4.33$\pm$0.63 & 0.85 & 0.89 \\
60  & 3.66$\pm$0.51 & 0.84 & 0.91 \\
70  & 3.31$\pm$0.40 & 0.91 & 0.93 \\
80  & 3.08$\pm$0.31 & 0.93 & 0.94 \\
90  & 2.86$\pm$0.30 & 0.93 & 0.94 \\
100 & 2.67$\pm$0.24 & 0.94 & 0.95 \\ 
150 & 2.21$\pm$0.16 & 0.83 & 0.82 \\
200 & 2.01$\pm$0.14 & 0.86 & 0.87 \\
250 & 1.69$\pm$0.12 & 0.89 & 0.89 \\
300 & 1.58$\pm$0.09 & 0.94 & 0.91 \\
350 & 1.46$\pm$0.11 & 0.92 & 0.93 \\
400 & 1.40$\pm$0.09 & 0.95 & 0.94 \\
500 & 1.28$\pm$0.10 & 0.91 & 0.89 \\
\enddata
\end{deluxetable}

\end{document}